\documentclass[aps,prd,preprint,groupedaddress]{revtex4-1}
\usepackage{setspace}
\usepackage{anysize}
\usepackage{graphicx}
\usepackage{latexsym}
\usepackage{amsmath}
\usepackage{amsfonts}
\usepackage{amssymb}
\usepackage[utf8x]{inputenc}
\usepackage{mathrsfs}
\usepackage{verbatim}

\newcommand{\bo}{\raise-1mm\hbox{\Large$\Box$}}
\newcommand{\expva}[1]{\langle #1 \rangle}

\begin{document}



\author{Saulo Diles}
\affiliation{Campus Salinópolis, Universidade Federal do Pará,
68721-000, Salinópolis, Pará, Brazil}
\affiliation{Laboratório de Astrofísica
Teórica e Observacional, Departamento de Ciências Exatas e Tecnológicas,
Universidade Estadual de Santa Cruz, 45650-000, Ilhéus, Bahia, Brazil}

 

\title{ The role of Weyl symmetry in hydrodynamics}



\begin{abstract}

This article is dedicated to the analysis of  Weyl symmetry in the context of relativistic hydrodynamics.
Here is discussed how this symmetry is properly implemented using the prescription of minimal coupling:  $\partial\to \partial +\omega \mathcal{A}$. 
It is shown that this prescription has no problem to deal with curvature since it gives the correct expressions for the commutator of covariant derivatives.

In hydrodynamics, Weyl gauge connection emerges from the  degrees of freedom of the fluid: it is a combination of
the expansion and entropy gradient. The remaining degrees of freedom, shear, vorticity and  the metric tensor, are see in this context as charged fields under 
the Weyl gauge connection. The gauge  nature of the connection provides natural dynamics to it via equations of motion analogous to the
Maxwell equations for electromagnetism. As a consequence, a charge for the Weyl connection is defined and the notion of local charge is analyzed generating the conservation 
law for the Weyl charge.

\end{abstract}

\keywords{Weyl Symmetry, General Relativity, Gauge Theory, Hydrodynamics}

\maketitle

\section{Introduction }  
The  holographic calculation of the $\eta/s$ ratio in the $\mathcal{N}=4$ SYM plasma  and the production of
the Quark-Gluon plasma at RICH and LHC reveled the relativistic nature of this fluid and the scaling symmetry of the system \cite{Policastro:2002se,Song:2012ua}. 
The trace of Weyl symmetry comes either from the experimental bound on bulk viscosity and from the similarity of the $\eta/s$ calculation
in AdS/CFT to the experimental data. This phenomenology calls attention to the importance of the dynamics of a relativistic fluid with  Weyl invariance.
 The formulation of the  relativistic fluid dynamics is performed in a gradient expansion approach 
 \cite{landau:59,Baier:2007ix,Bhattacharyya:2008mz,Banerjee:2008th,Heller:2014wfa, Grozdanov:2015kqa}.
 The elastic and friction properties of the fluid are described as perturbations of the ideal fluid flow and the dispersion relation is expressed 
 in a power series $\omega(k)= \omega_0 +\delta_1k +\delta_2k^2 +...$ Perturbations are considered at the  level of energy-momentum tensor, starting with
 ideal fluid
\begin{equation}
T_{_{ideal}}^{a b}= \epsilon u^au^b + p\Delta^{ab},
\end{equation}
with $\epsilon$ the  energy density, $p$ the pressure, $u^a$ a normalized  velocity field ($g_{ab}u^au^b=-1$) and $\Delta^{ab} = g^{ab} + u^au^b$ is the 
transverse projector. The signature of the metric is $diag(g)=(-,+,...,+)$. The  gradient expansion of energy momentum-tensor is 
\begin{equation}
T^{ab} = (\epsilon +  E ) u^au^b +(p +  P )\Delta^{ab} + q^{(a}u^{b)} + \pi^{\expva{ab}},
\end{equation}
that includes scalar $(E,P)$, vector $(q^a)$ and tensor $\pi^{\expva{ab}}$ perturbations of the ideal fluid. 
The symmetric, transverse and traceless part of a contra-variant rank 2 tensor is
$  \mathcal{T}^{\expva{ab}}= \frac{1}{2}\Delta^a_{~c}\Delta^b_{~d}(\mathcal{T}^{cd}+\mathcal{T}^{dc}) - \frac{1}{d-1} \Delta^{ab}\Delta_{cd}\mathcal{T}^{cd} .$
These perturbations arise from the complete set of hydrodynamical degrees of freedom: the fluid entropy $\nabla_{\perp}^a \ln s$, the velocity 
$\nabla_{_\perp a} u^b$ and the torsion free connection $\Gamma_{~ab}^c = \frac{1}{2}g^{ck}(\partial_a g_{bk} +\partial_bg_{ka} - \partial_k g_{ab})$ 
presents in the covariant derivatives and in the curvature tensors. Velocity gradient is separated by symmetry, we have
expansion $\Theta = \nabla_{_\perp a} u ^a$,  shear $\sigma^{ab} \equiv 2\nabla_{_\perp}^{\langle a}u^{b\rangle}$ and vorticity 
$\Omega^{ab}\equiv \nabla_{_\perp}^{[a}u^{b]}$. 

The ideal fluid equations together with the thermodynamical equations fixes that 
$\nabla_{_\perp}^a\ln s = -(d-1) u^c\nabla_c u^a ,~~~u^a\nabla_a \ln s= -\Theta,$ 
reducing the effective degrees of freedom. Gradient expansion is them completely fixed by  shear, vorticity, expansion, entropy gradient  $\nabla_{_\perp}^a\ln s$, 
Riemann tensor and their derivatives. The number of derivatives in each structure is the order of gradient expansion where it appears and for each structure in this 
expansion there is one transport coefficient associated. Weyl symmetry constrains the dynamics of the fluid, reducing the number of transport coefficients in all orders. 
When this symmetry takes place expansion and entropy gradient are no longer allowed by symmetry in the gradient expansion, instead they combine forming the gauge 
connection. It is shown that minimal coupling correctly realizes Weyl invariance even  when it comes to curvature structures. 
 The gauge structure of Weyl symmetry is explored in the context of relativistic hydrodynamics, revealing that
a constraint to its dynamics due to a conservation law associated with the Weyl gauge charge.

This article is organized as follows: section 2 is dedicated to analyze Weyl symmetry in a general scenario and the minimal coupling prescription is established,
in section 3 is discussed the consequences of this symmetry in a hydrodynamical system, section 4 deals with the notion of local charge
conservation for the Weyl gauge field and in section 5 are made some final comments.

\section{Minimal coupling prescription} 
To require Weyl symmetry we  impose that the system is invariant under the local scaling transformation 
\begin{equation}
g^{ab} \to e^{-2\phi(x)}g^{ab},~~ g_{ab} \to e^{2\phi(x)}g_{ab}.
\end{equation}
We say that $g^{ab}$ has scaling weight $2$ while $g_{ab}$ has scaling weight $-2$.
The local scaling of the metric requires a scaling of the velocity field $u^a\to e^{-\phi(x)}u^a.$ 
Weyl symmetry is a local scaling invariance which in some cases is equivalent to conformal symmetry \cite{Iorio:1996ad,Jackiw:2011vz}.
For an hydrodynamic system to be invariant under such a transformation its energy-momentum tensor should 
transform as a tensorial density with scaling weight $\omega_{T}=d+2$:
\begin{equation}
 T^{ab} \to e^{-(d+2)\phi}T^{ab}.
\end{equation}
Consequently, in the gradient expansion only Weyl covariant perturbations are allowed. 

A non-trivial role of the Weyl scaling is that it changes the Christoffel connection in a inhomogeneous way:
$$\Gamma_{~ab}^c\to \Gamma_{~ab}^c +\delta^c_{~a}\nabla_b\phi + \delta^c_{~b}\nabla_a\phi - g_{ab}\nabla^c \phi.$$
As a consequence the velocity gradients and the curvature tensor also transforms inhomogeneously under Weyl transformations. To repair the inhomogeneous
terms of the hydrodynamical degrees of freedom it was  introduced in \cite{Loganayagam:2008is} a Weyl covariant derivative that acts in the fields preserving their character
of tensor density. Take a vector field $\zeta^b$ with scaling weight $\omega_\zeta,~~\zeta^b\to e^{-\omega_{\zeta}\phi}\zeta^b,$ its Weyl covariant derivative is  
\begin{equation}
 \mathcal{D}_a\zeta^b = \nabla_a\zeta^b + \omega_\zeta \mathcal{A}_a\zeta^b + (g_{ac}\mathcal{A}^b -\delta^b_{~a}\mathcal{A}_c-\delta^b_{~c}\mathcal{A}_a)\zeta^c,  
\end{equation}
where $\mathcal{A}$ is the Weyl connection and transform as $\mathcal{A}_a\to \mathcal{A}_a +\partial_a\phi$ under Weyl transformations. 
The mathematical properties of this connection are discussed in \cite{hall:92}.
This structure of the Weyl covariant derivative can be organized in a subtle way. It is useful to look at Weyl covariant derivative using
the explicit form of the Christoffel connection:
\begin{align}
\mathcal{D}_a\zeta^b =~ &\partial_a\zeta^b + \omega_\zeta \mathcal{A}_a\zeta^b  +  \nonumber \\ 
\frac{1}{2}g^{bk}(\partial_c g_{ak} +\partial_ag_{kc} - \partial_k g_{ca})\zeta^c &+ (g_{ac}\mathcal{A}^b -\delta^b_{~a}\mathcal{A}_c-\delta^b_{~c}\mathcal{A}_a)\zeta^c. 
\end{align}
We can write the first two terms in the nice composition $(\partial_a + \omega_\zeta A_a)\zeta^b$. 
Moreover, we can proceed in the same way and identify the terms of the Christoffel connection with the terms of the Weyl gauge connection in the simple form 
\begin{equation}
 \mathcal{D}_a\zeta^b = (\partial_a + \omega_\zeta\mathcal{A}_a)\zeta^b +   \frac{1}{2}g^{bk}[(\partial_c+\omega_g\mathcal{A}_c) g_{ak} +
 (\partial_a+\omega_g\mathcal{A}_a)g_{kc} - (\partial_k + \omega_g\mathcal{A}_k) g_{ca})]\zeta^c.  
\end{equation}
For tensors with rank bigger the one we have one combination of the  connections for each Christoffel connection such that we can always combine
one metric derivative with one Weyl connection as  expressed above for the contra-variant vector.
It means that in order to implement Weyl covariance we can just replace the partial derivative $\partial_a$ by the Weyl invariant one:
\begin{equation}
\partial_a \mathcal{O} \to (\partial_a   + \omega_\mathcal{O} \mathcal{A}_a)\mathcal{O}. \label{replacement}
\end{equation}
In this approach the invariance under local scaling is replaced by the manifestation of a one-form gauge field $\mathcal{A} = \mathcal{A}_adx^a$.
This gauge field does not represent physical degrees of freedom since it is a connection, the corresponding observables are the components of the gauge invariant tensor 
\begin{equation}
 \mathcal{F}_{ab} = \nabla_a \mathcal{A}_b -\nabla_b \mathcal{A}_a = \partial_a \mathcal{A}_b - \partial_b \mathcal{A}_a .
\end{equation}

\subsection{Curvature}
To show how this replacement works  the commutation of the Weyl-covariant derivatives is calculated using this approach and will be shown that it provides correctly
the well known result
\begin{equation}
[\mathcal{D}_a,\mathcal{D}_b]\zeta^c = \omega_\zeta\mathcal{F}_{ab}\zeta^c + \mathcal{R}_{ab~~d}^{~~~c}\zeta^d, \label{curvature}
\end{equation}
where $\mathcal{R}_{ab~~d}^{~~~c}$ is the conformal Riemann tensor defined by
\begin{align}
 \mathcal{R}_{ab~~d}^{~~~c} = R_{ab~~d}^{~~~c} + \nabla_a[ \delta_b^{~c}\mathcal{A}_d + \delta_d^{~c}\mathcal{A}_b - g_{bd}\mathcal{A}^c ] 
 &- \nabla_b[\delta_a^{~c}\mathcal{A}_d + \delta^{~c}_d\mathcal{A}_a - g_{ad}\mathcal{A}^c ]  \nonumber \\
-[\delta_a^{~f}\mathcal{A}_d + \delta^f_{~d}\mathcal{A}_a - g_{ad}\mathcal{A}^f][&\delta_b^{~c}\mathcal{A}_f + \delta^c_{~f}\mathcal{A}_b -g_{bf}\mathcal{A}^c] \nonumber \\
+[\delta_b^{~f}\mathcal{A}_d+ \delta^f_{~d}\mathcal{A}_b -g_{bd}\mathcal{A}^f ][&\delta_a^{~c}\mathcal{A}_f + \delta^c_{~f}\mathcal{A}_a -g_{af}\mathcal{A}^c].\label{riem}
\end{align}
First note that the above equation can be obtained using the above prescription of minimal coupling.
It is straightforward to show that the conformal Riemann tensor $ \mathcal{R}_{ab~~d}^{~~~c}$ is obtained by taking
the usual Riemann tensor expressed in terms of the derivatives of the metric field $\partial g$ and performing the replacement of 
$\partial g$ by $(\partial + \omega_g \mathcal{A})g$. 

It is useful to look at the usual Christoffel connection as a vector operator $\Gamma_a$. This operator can act in arbitrary tensor,
its action is defined as follows: for a contra-variant vector
$\Gamma_a\zeta^b = \Gamma_{af}^b\zeta^f$, for a covariant vector $\Gamma_a\alpha_b = -\Gamma_{ab}^f\alpha_f$, for a rank 2 tensor
$\Gamma_aG^b_{~c} =  \Gamma_{ak}^bG^k_{~c} -\Gamma_{ac}^kG^b_{~k}$ and so on. With this definition the geometrical covariant derivative takes the nice form
$\nabla_a = \partial_a + \Gamma_a$. The usual Riemann tensor is defined by the action of the commutator of the diffeomorphic covariant derivative  on a vector field
$[\nabla_a,\nabla_b]\varphi^c = R_{ab~~d}^{~~~c}\varphi^d$ and can be expressed as follows
\begin{equation}
 [\nabla_a,\nabla_b]\zeta^c = [\partial_a + \Gamma_a, \partial_b + \Gamma_b]\zeta^c = \left([\partial_a , \Gamma_b] + [ \Gamma_a, \partial_b ]
 + [\Gamma_a,\Gamma_b]\right)\zeta^c = R_{ab~~d}^{~~~c}\zeta^d. \nonumber
\end{equation}
Apply the minimal coupling prescription means that the replacement 
$\partial_a\zeta \to (\partial + \omega_\zeta \mathcal{A})\zeta$ and $\Gamma[\partial g]\to \bar{\Gamma}=\Gamma[(\partial+\omega_g\mathcal{A})g]$ 
is performed and one find that 
\begin{equation}
  [\mathcal{D}_a,\mathcal{D}_b]\zeta^c = \omega_\zeta([\partial_a , \mathcal{A}_b]+[\partial_b,\mathcal{A}_a])\zeta^c + \left([\partial_a , \bar{\Gamma}_b] + 
  [ \bar{\Gamma}_a, \partial_b ] + [\bar{\Gamma}_a,\bar{\Gamma}_b]\right)\zeta^c.
\end{equation}
In first bracket appear the curvature for the gauge connection $\mathcal{F}_{ab} = [\partial_a , \mathcal{A}_b]+[\partial_b,\mathcal{A}_a]$. 
The second term on the hight hand side is the Riemann tensor calculated using the new Christoffel symbol $\bar{\Gamma}$, 
where the metric is minimally coupled to the Weyl connection $\mathcal{A}$, and  is precisely the conformal Riemann tensor
$\left([\partial_a , \bar{\Gamma}_b] +   [ \bar{\Gamma}_a, \partial_b ] + [\bar{\Gamma}_a,\bar{\Gamma}_b]\right)\zeta^c=\mathcal{R}_{ab~~d}^{~~~c}\zeta^d$. 
The commutator of covariant derivatives acting on the two-form field $\mathcal{F}$ is
\begin{equation}
 [\mathcal{D}_a,\mathcal{D}_b]\mathcal{F}_{cd} = -\mathcal{R}_{ab~c}^{~~i}\mathcal{F}_{id}-\mathcal{R}_{ab~d}^{~~i}\mathcal{F}_{ci}. \label{curvatures}
\end{equation}
The Weyl covariant version of the Riemann tensor has different symmetry properties as pointed out in  \cite{Loganayagam:2008is}.
The new Ricci tensor $\mathcal{R}_{ab} = \mathcal{R}_{ac~~b}^{~~~c}$ is no more symmetric, $\mathcal{R}_{[ab]} = -d \mathcal{F}_{ab},$
as a consequence of the fact that the partial derivative commute while the gauge covariant one do not.

This approach differs in point of view from reference \cite{Iorio:1996ad}, where is said that the Ricci tensor 
is coupled to the Weyl connection through a non-minimal coupling. What is shown here is that if one be careful to implement Weyl covariance through minimal coupling it will
either generates the correct gauge curvature $\mathcal{F}$ and also correctly change the geometric curvature $R\to \mathcal{R}$ giving the eq.(\ref{curvature}).
The present analysis provides a framework to implement Weyl invariance using only what the authors of ref.\cite{Iorio:1996ad} calls \textit{Weyl gauging}. 

\section{Hydrodynamics and Weyl invariance}
In hydrodynamics the gradient expansion is build up by correcting the energy-momentum  tensor of an ideal fluid using derivatives of the fundamental degrees of freedom 
combined in structures allowed by the underlying symmetries of the system. In this sense Weyl symmetry plays two fundamental roles in such a system. 
At the first place it constrains the corrections of the energy-momentum tensor to be Weyl invariants, reducing the effective number of transport coefficients in all orders
of gradient expansion. This mechanism is discussed along this section. In a second place the Weyl symmetry constrains the dynamics of the system via 
a local charge conservation law, which is presented in the next section. This local charge conservation is expected as far as  Weyl symmetry is treated here as a local gauge
one. At the quantum level, the requirement of trace anomaly cancellation provides additional constrains in the fourth order transport coefficients, 
as pointed out in \cite{Grozdanov:2015kqa}.

We look at the gradients of the hydrodynamical degrees of freedom. The starting point are the first order gradients 
$\Theta, \sigma^{ab},\Omega^{ab},\nabla_{\perp a}\ln s$ and the Riemann tensor  $R_{abcd}$ which is second order. One can note that
shear and vorticity  are already  Weyl covariant  and can be expressed using the Weyl covariant derivative: 
$ \sigma^{ab}=\mathcal{D}^au^b+\mathcal{D}^bu^a,~~\Omega^{ab}=\mathcal{D}^au^b-\mathcal{D}^bu^a.$
Expansion and entropy gradient do not transform covariantly under Weyl scaling, instead they combine into the  connection for  Weyl gauge symmetry:
\begin{equation}
 \mathcal{A}_a\equiv \frac{1}{d-1}[\nabla_{\perp a}\ln s - \Theta u_a],~~~\mathcal{A}_a\to \mathcal{A}_a + \partial_a\phi. \label{connection}
\end{equation}
In a gauge theory the connection is not a physical observable since it is not gauge invariant. The relevant structure is the Maxwell tensor field 
$\mathcal{F}_{ab}=\partial_a \mathcal{A}_b - \partial_b \mathcal{A}_a,$ whose components are second order in derivatives.   
Due to its antisymmetric nature, it can appear at lowest order  in gradient expansion contracted with shear and vorticity 
generating tensors that are third order in derivatives: $\mathcal{F}_c^{~~\langle{a}}\sigma^{b \rangle c},~~\mathcal{F}_c^{~~\langle{a}}\omega^{b \rangle c}$.
A Lagrangean density for $\mathcal{A}$ is possible and is a fourth order scalar $\mathcal{F}^2 = 2\mathcal{E}^2-2\mathcal{B}^2$ with a corresponding 
 energy-momentum tensor that is also fourth order: $T_{ab}=g_{ab}\mathcal{F}^2 + \mathcal{F}_{ac}\mathcal{F}_b^{~c}.$ 

The curvature structures in this context are given by contractions of the conformal Riemann tensor $\mathcal{R}_{abcd}$ with metric and velocity field.
The structures obtained in this way completely replaces the corrections combining the usual Riemann tensor $R_{abcd}$ with the remaining 
degrees of freedom usually referred in higher order hydrodynamics.

This is the way that Weyl symmetry constrains the number of transport coefficients, it maps the set of  fundamental structures 
used to build up the gradient expansion of a non-conformal fluid into the smaller set of fundamental structures compatible with Weyl symmetry:
$\{\sigma^{ab},\Omega^{ab},\Theta,\nabla_{\perp}^a ln s, R_{abcd}\}\to \{\sigma^{ab},\Omega^{ab},\mathcal{F}_{ab}, \mathcal{R}_{abcd}\}.$
So, in first order the absence of bulk viscosity is the signal of Weyl invariance. The effects of the gauge curvature are of second order,
even that the gauge curvature do not appear at this level in the energy momentum tensor, it will appear first time in gradient expansion at third order.
At fourth order we have  the action of the gauge field and the associated energy-momentum tensor. 
 
\subsection{Electric and Magnetic components of the Weyl gauge field}

The electric and magnetic components of the field strength $\mathcal{F}_{ab}$ are 
\begin{equation}
 (\vec{\mathcal{E}})_i = \mathcal{F}_{0i},~~( \vec{\mathcal{B}})_i = \frac{1}{2}\epsilon_i^{~jk}\mathcal{F}_{jk},~~i,j,k = 1,2,...,d-1.
\end{equation}
We consider as an illustrative example the electric and magnetic fields observed in the  in the special case where space-time is flat and take a local rest frame 
where $u_a=(-1,\vec{0})$. Them the transverse derivative becomes a space gradient $\nabla_{_\perp}^a = (0,\vec{\nabla})$ and the gauge connection is
$\mathcal{A}_a = (0, \vec{\nabla}ln s)$ so that 
\begin{equation}
(\vec{\mathcal{E}})_i =  -\frac{\partial_i(\partial_t ln s)}{d-1},~~ (\vec{\mathcal{B}})_i =  -\frac{\epsilon_i^{~jk}\partial_j\partial_kln s}{d-1} = 0. 
\end{equation}
One can go further in analogy with classical electrodynamics and speculate about a Newton force
that the  fluid exerts in a test particle. If the test particle has scaling dimension $\omega_p$ one expect, in analogy 
with Lorentz force, that this particle is accelerated by the force $ \vec{\mathbf{F}}_{\omega_p} =\omega_p \vec{\mathcal{E}}. $
This expression is a speculation about the effect of an hydrodynamical configuration in a classical test particle.
The notion of  Weyl charge is not so clear for a classical particle, all the time we have been talking about classical fields with well defined behavior under
local scaling. If we have a particle described by a quantum field its scaling dimension is fixed by the nature of this quantum field so the quantum nature of the
particle may determine its scaling dimension. 

The ``electromagnetic" structure of Weyl symmetry was noted from the early times that Weyl invariance was proposed \cite{Weyl:1918ib} and
its  minimal coupling mechanism  was first discussed  some years later  \cite{London:1927fk}. At this time the community tried to look at the Weyl symmetry
as responsible for generating the electromagnetic field of Maxwell theory, but this idea was set aside and one important reason was
the difficult to realize the real factor $\omega_g$ as an imaginary electric charge of the metric. 
In contrast, when one analyses the Weyl symmetry in hydrodynamics it is clear that neither the scaling weight is an imaginary electric charge 
nor the Weyl symmetry generates Maxwell field. The requirement of Weyl covariance in this approach means that expansion and entropy gradient of an arbitrary fluid
manifests as a gauge connection. In this sense the metric field $g_{ab}$ is charged under this gauge  connection with charge $\omega_g=-2$, 
in the same way the shear and vorticity matrices $\sigma_a^{~b},~~\Omega_a^{~b}$ are charged with charge $\omega_\sigma=\omega_\Omega=+1$. 
It is important to emphasize that Weyl connection differ from the electromagnetic connection since
some articles, for example in ref.\cite{Cabral:2016yxh},  refer to them as being equivalent. 

\subsection{Two approaches for Weyl symmetry}
The conformal Riemann tensor defined in eq.(\ref{riem}) is Weyl invariant and depends explicitly on the gauge connection, 
$\mathcal{R}=\mathcal{R}(g,\mathcal{A}).$ In contrast the Weyl tensor $C_{abcd}$
is build up from combinations of the Riemann tensor with the metric and do not depend on the Weyl gauge connection, $C=C(g)$.
In order to implement Weyl symmetry it is sufficient to introduce a gauge connection and a minimal coupling prescription, but it is not a necessary condition.
A Weyl invariant gravity theory for example is obtained by considering combinations of the usual curvature tensors, see  \cite{Zee:1981ff,Flanagan:2006ra}.
Also for analyzing a hydrodynamical system one can  define combinations of the velocity field and curvatures tensor that are Weyl invariant, 
it is the so called \textit{gravito-electromagnetism} \cite{Maartens:1996ch,Maartens:1997fg}. In \textit{gravito-electromagnetism} the projections of the Weyl tensor
$C_{abcd}$ in the direction of a congruence $u^a$ are identified as the electric and magnetic tensors $E_{ab} = C_{acbd}u^cu^d,~~H_{ab} = {^*E}_{ab}$. 

There are two ways to deal with Weyl symmetry: one is to look for combinations of the fundamental structures of the theory that are Weyl invariants
and the other is to define a gauge connection and perform minimal coupling prescription. It is not required a manifestation of the gauge field $\mathcal{A}$
to reach Weyl invariance, but once there is a vector field in the theory that plays the role of a connection them the minimal coupling prescription guarantee Weyl 
invariance. So, in the description of Weyl invariant hydrodynamics the combination of expansion and gradient entropy into a vector field that changes by a total gradient 
under local Weyl scaling, eq.(\ref{connection}), tells us that the nature of an hydrodynamical system choose this way, as an abelian gauge theory, 
to manifest Weyl symmetry.

\section{Local charge conservation}
 
 Once we have a tensor field for the connection we can look for a global charge in the sense of an electric  Gauss law. 
For a given $d$ dimensional hyper volume $\Sigma$ the gauge charge inside it is given by the flux of the electric components of $\mathcal{F}$ through its boundary 
$\partial\Sigma$ 
\begin{equation}
 Q_{_{Weyl}} = \int_{\partial \Sigma} dA_i\mathcal{F}^{0 i}. \label{weylcharge}
\end{equation}
 In flat space it takes the familiar form of Gauss law $Q_{_{Weyl}} = \int_{\partial \Sigma} dA \hat{n}\cdot \vec{\mathcal{E}}.$
 
Equation (\ref{weylcharge}) defines a total amount of Weyl charge
and one can ask about the origin of it. We know that Coulomb charge is quantized in the atomic scale, it is carried only by electrons and protons 
with individual charges $-e,~+e$ respectively, so that the total amount of Coulomb charge inside a volume is the balance of  positive  and  
negative charge carries. A natural question arises in the context of Weyl gauge theory: Is the charge defined by eq.(\ref{weylcharge})  a total amount of  quantized 
charges carried by some particle-like excitation?

In classical electrodynamics the local notion of charge and charge flux in a reference frame are fixed by the current $J^\mu_e = \gamma(\rho_e,\rho_e\vec{v}).$
Maxwell theory has a Lageangean density  $\mathcal{L}_{_{Maxwell}}= -1/4F_{\mu\nu}F^{\mu\nu} + A_\mu J^\mu_e$ which extremization gives the equations of motion
$\partial_\nu F^{\mu\nu} = J^{\mu}_e$. The current $J_e$ is the source of the Maxwell field $F$. 
Anti-symmetry of $F^{\mu\nu}$ ensures that $\partial_\mu J^{\mu}_e = [\partial_\mu,\partial_\nu]F^{\mu\nu}=0$, which
is the continuity equation for the local conservation of electric charge. 

The electric Gauss law tells us that a local accumulation of charge generates
divergence of the electric field $\rho_e = \vec{\nabla}\cdot\vec{E}.$ 
So, in analogy with Maxwell theory the notion o local  Weyl charge should be given by the Gauss law for its ``electric"  field
$\mathcal{E}_i = \mathcal{F}_{0i}.$ The idea is that if we want to find a charge conservation law for the   Weyl charge we need to define
the notion of local charge density to build up a current  which sources the connection.
Consider for simplicity a flat space-time, the Gauss law for a bounded hyper-volume $\Sigma$ with boundary  $\partial \Sigma$ defines the local density 
of Weyl charge   as follows
\begin{equation}
 Q_{Weyl} \equiv \int_{\partial \Sigma}\vec{\mathcal{E}}\cdot d\vec{A} = \int_{\Sigma} (\vec{\nabla}\cdot\vec{\mathcal{E}})dV =: \int_{\Sigma}\rho_w dV,
\end{equation}
and  the Weyl density charge is $ \rho_w(x) \equiv \vec{\nabla}\cdot\vec{\mathcal{E}},$ in curved background we have that $\rho_w\sim \nabla_i\mathcal{F}^{0i}.$
The proposal is that the associated conserved current 
is defined in the rest frame by $\mathcal{J}_{_{rest}}^a = (\rho_w,\vec{0}),$ so that in a boosted frame  $\mathcal{J}^a = ( \gamma\rho_w, \gamma\rho_w \vec{v}).$
The current $\mathcal{J}^a$ will sources the tensor field $\mathcal{F}$, providing an action for the Weyl connection: 
$\mathcal{S}[\mathcal{A}] = c_w\int dV \sqrt{-g} \left( -\frac{1}{4} \mathcal{F}_{ab}\mathcal{F}^{ab} + \mathcal{A}_a\mathcal{J}^a\right), $
extremizing this action above with respect to $\mathcal{A}$ results in the equation of motion 
\begin{equation}
 \mathcal{D}_b\mathcal{F}^{ab} = \mathcal{J}^a. \label{weyldynamics}
\end{equation}
In a rest frame and in flat space-time we have  $\mathcal{J}^0_{_{rest}} = \mathcal{D}_i\mathcal{F}^{0i} = \vec{\nabla}\cdot\vec{\mathcal{E}},$ as expected.
Taking the divergence of the eq.(\ref{weyldynamics}) gives  $[\mathcal{D}_a,\mathcal{D}_b]\mathcal{F}^{ab} = \mathcal{D}_a\mathcal{J}^a$. Using $\mathcal{D}_ag^{bc} = 0$ 
together with eq.(\ref{curvatures}) we find 
\begin{equation}
 \mathcal{D}_a\mathcal{J}^a = 0, \label{conservation}
\end{equation}
which is the local conservation of the Weyl gauge current. This charge conservation is intriguing, we know that the gauge nature of Weyl symmetry
provides a divergence-less vector field that is analogous to the Maxwell electric current. But in classical electrodynamics  local charge density is 
define by $\rho_e(\vec{r},t) = e(\sum_j\delta^3[\vec{r} - \vec{r}^{~p}_j(t)] - \sum_i\delta^3[\vec{r} - \vec{r}^{~e}_i(t)]),$ 
where $\vec{r}^{~p}_j(t)$ and $\vec{r}^{~e}_i(t)$
are respectively the position of the j-th proton and of the i-th electron. So in classical electrodynamics the charge conservation constrains 
the trajectories of the charge carries by providing one additional coupled differential equation involving all positions $ \vec{r}^{~p}_j(t),~~\vec{r}^{~e}_i(t)$
of the point particles in the system.

For  Weyl invariant hydrodynamics the notion of charge carrier is absent
and it is not evident the physical meaning of the charge conservation in eq.(\ref{conservation}) in therms of classical trajectories of point particles. 
Even so eq.(\ref{conservation}) constrains the dynamics of an hydrodynamical system, since the conserved current is a function of its degrees of freedom.
This constrain arises from the assumption that the tensor field $\mathcal{F}$ should be sourced by the current $\mathcal{J}$ in such a way that eq.(\ref{weyldynamics})
is satisfied. In the point of view of the dynamics of a gauge field it seems completely natural. By the other hand, it is absolutely non-trivial for hydrodynamics
where one deals with an effective description. It is the gauge nature of the Weyl symmetry that provides this additional restriction in conformal hydrodynamics.

\section{Final Comments} 
The consideration of minimal coupling to implement Weyl invariance appears as a powerful tool to deal with conformal hydrodynamics. 
The gauge nature of Weyl symmetry reveals a very close analogy with the $U(1)$ gauge symmetry of electrodynamics but its gauge group is not compact
and we cannot interpret the Weyl charge as an electromagnetic charge. 

The analysis of curvatures structure revealed the power of minimal coupling approach. 
There is no need to define other gauge transformations but the Weyl scaling, the role for the gauge connection is enough. For a Weyl invariant fluid the connection
is given by expansion and entropy gradient, which get  their dynamics constrained and loose their status of physical observables when Weyl symmetry takes place. 
The gauge nature of Weyl symmetry also leads to the emergence of a local charge conservation. The notion of charge in this gauge theory is not clear at the fundamental
level since there is no known Weyl-charge carrier. 

 It is known that the vanishing of the shear tensor requires that the fluid is expansion free or  vorticity-free \cite{glass75,collins86,Senovilla:1997bw}. 
 In the second case the remaining physical degrees of freedom are the components $\vec{\mathcal{E}},\vec{\mathcal{B}}$
 of the stress tensor and the space-time geometry $g$. The approach presented here can be used as a guide in the search of a Lagrangean formulation for these
 hydrodynamical systems.

\acknowledgments{ I thank to Alex S. Miranda for important discussions on relativistic hydrodynamics and for the good reception at UESC.
I also thank to Amilcar Queiroz for important suggestions and good discussions. I thank to Camila Aranha for the company all this time.
This work was supported by the PNPD  program of CAPES. }

 \end{document}